# KAMLAND-EXPERIMENT AND SOLITON-LIKE NUCLEAR GEOREACTOR

*V.D.Rusov[1], D.A. Litvinov[1], S. Ch. Mavrodiev[2], E.P. Linnik[1], V.N. Vaschenko[3],*

*T.N. Zelentsova[1], V.A. Tarasov[1]*

[1]*Odessa National Polytechnic University, Ukraine*,
[2]*Institute for Nuclear Research and Nuclear Energy, BAS, Sofia, Bulgaria*
[3]*National Antarctic Scientific Centre, Kiev, Ukraine*

## Abstract

We give an alternative description of the new data produced in the KamLAND experiment, assuming the existence of a natural nuclear reactor on the boundary of the liquid and solid phases of the Earth's core. Analyzing the uncertainty of antineutrino spectrum of georeactor origin, we show that the theoretical (which takes into account the soliton-like nuclear georeactor with power about 20 TW) reactor antineutrino spectrum describes with good accuracy the new experimental KamLAND-data. At the same time the parameters of mixing ($\Delta m^2_{21}$=2.5·10⁻⁵ eV², tan²$\theta_{12}$=0.437) calculated within the framework of georeactor hypothesis are substantially closer to the data of solar flux SNO-experiment then the parameters of mixing obtained in KamLAND-experiment.

Today it is obvious that the experiments of KamLAND-collobaration which are conducted during last five years [1-4] are extremely important not only for the observation of reactor antineutrino oscillations, but they make it possible to verify for the first time one of the most vivid and mysterious ideas in nuclear geophysics – the hypothesis of the existence of natural nuclear georeactor [5]. In spite of its singularity and long history, today this hypothesis is especially attractive because it enables clearly to explain in the view of physics many, on the face of it, unrelated geophysical anomalous effects, whose fundamental character is beyond any doubt.

First of all it concerns the problem of ³He isotope origin in the Earth interior, whose concentration, as is well known, "mysticly" increases to the center of Earth. This is practically impossible to explain within the framework of classic geophysical notions, because, on the one hand, the existence of this isotope in nature is caused and predetermined exclusively by nuclear processes, and, on the other hand, the high mobility and chemical inertness of $^3$He atoms exclude their retention by any chemical or physical traps. For example, as is customary to consider within the framework of hypothesis of "solar" helium in the Earth's core, which was captured as far back as the period of accretion of the Earth and slowly seeped to the surface during a few milliards years.

_________________________________________________________________

[*] Corresponding author: Prof. Rusov V.D., E-mail: siiis@te.net.ua

A potent argument in favour of nuclear georeactor existence are results of recent seismo-tomography researches of abnormally high global core-mantle boundary heat flow (13±4 TW), which is several times higher than the value of radiogenic heat in lower mantle ($D$"-region) [7]. To explain such an anomalous high heat flow the authors of this paper advanced the hypothesis of young age for the Earth's solid core, where the crystallization energy of the young solid core is the reason of anomalous temperature effect.

In full measure it concerns the known problem of the nature of power source maintaining the convection in the Earth's liquid core or, more precisely, the mechanism of magneto-hydrodynamic dynamo generating the Earth's magnetic field. Obviously, that the well-known $^{40}$K-mechanism of radiogenic heat production in the solid core of the Earth does not solve a problem in the whole, because it can not explain the heat flow energy balance on a core-mantle boundary. It should be noted also the so-called mechanism of the Earth's magnetic field inversions closely associated with the problem of convection in a liquid core. Surprisingly, both these fundamental mechanisms have simple and clear physical explanation within the framework of hypothesis of the existence of natural nuclear georeactor on the boundary of the liquid and solid phases of the Earth [5,6] too.

If the georeactor hypothesis is true, the fluctuations of georeactor thermal power can influence on the Earth's global climate as anomalous temperature jumps due to following circumstances. Strong fluctuations of georeactor thermal power can lead to the partial blocking of convection in a liquid core [6] and to the change of angular velocity of liquid geosphere rotation, and in that way by virtue of conservation law of Earth's angular moment to the change of angular velocity of mantle and the Earth's surface, respectively. This means that heat or, more precisely, dissipation energy caused by the friction of earthly surface and bottom layer can make a considerable contribution to atmosphere general energy balance and thereby substantially to influence on the Earth's global climate evolution [6].

However, in spite of obvious efficiency and allurement of hypothesis of natural nuclear georeactor existence, the main difficulties of its perception are predetermined by non-trivial properties which georeactor must possess. At first, the natural, i.e. unenriched uranium or thorium must be used as a nuclear fuel. Secondly, the reactivity regulation system of reactor by traditional control rods is completely absents, but for all that a reactor must possess the property of so-called internal safety. It means that the critical state of reactor core must be continually maintained in any situation, i.e. the normal operation of the reactor is automatically maintained not as a result of operator activity, but by virtue of physical reasons-laws preventing the explosive development of chain reaction by the natural way. Figuratively speaking, the reactor with internal safety it is "nuclear installation which never explode" [8].

Strangely enough, but reactors satisfying such an unusual requirements are possible in reality. For the first time the idea of such reactor was proposed by Feoktistov [9] and independently by Teller, Ishikawa and Wood [10].

The main idea of reactor with internal safety consists in the selection of fuel composition so that, at first, characteristic time $\tau_\beta$ of the nuclear burning of fuel active (fissile) component is substantially greater than the characteristic time of delayed neutrons production and, secondly, the necessary self-regulation conditions are meet during reactor operation (that always take place, when the equilibrium concentration of fuel active component is greater than critical concentration [9]). These very important conditions can practically always to be attained, if among other reactions in a reactor the chain of nuclear transformations of Feoktistov uranium-plutonium cycle type [9]

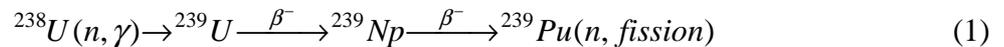

$$^{238}U(n,\gamma) \rightarrow {}^{239}U \xrightarrow{\beta^-} {}^{239}Np \xrightarrow{\beta^-} {}^{239}Pu(n, fission) \qquad (1)$$

or Teller-Ishikawa-Wood thorium-uranium cycle type [10]

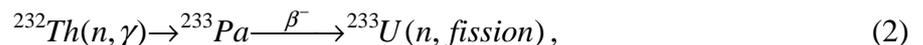

$$^{232}Th(n,\gamma) \rightarrow {}^{233}Pa \xrightarrow{\beta^-} {}^{233}U(n, fission), \qquad (2)$$

will be enough appreciable

In both cases the produced fissile isotopes of $^{239}$Pu or $^{233}$U are nuclear fuel active components. Characteristic time of such reaction, i.e. the time of proper $\beta$-decays, is approximately equal to $\tau_\beta$ =2.3/ln2≈3.3 days and $\tau_\beta$ ≈39.5 days for reactions (1) and (2), respectively, that is several orders greater than the time of delayed neutrons production.

Self-regulation of the nuclear burning process (at indicated above ration between equilibrium and critical concentrations of fuel active components [9]) is stipulated by the fact that such system left by itself can not pass from a critical condition to the reactor acceleration mode, because a critical concentration is bounded from above by finite value of plutonium equilibrium concentration, i.e. $\tilde{n}_{Pu} > n_{crit}$. On phenomenal level the self-regulation of nuclear burning is manifested as follows. The neutron flux increase due to some reasons will result in the rapid burnup, for example, of plutonium, i.e. to decrease of its concentration as well as the neutron flux decrease, while the new nuclei of $^{239}Pu$ are produced with the same generation rate during about $\tau_\beta =3.3$ days. And vice versa, if the neutron flux is sharply decreased due to external action, the burnup rate decrease too and plutonium accumulation rate will be increased as well as the number of neutron production in approximately same time. The analogical situation will be observed for thorium-uranium cycle (2), but in time $\tau_\beta =39.5$ days.

Generation of the system of kinetic equations for components of nuclear fuel and neutrons (as the diffusion approximation) in such chains is sufficiently simple and was described in detail in our paper [5]. The solutions as concentration soliton-like waves of nuclear fuel components and neutrons which are typical for such problem (see Eqs. (3)-(9) in Ref. [5]) for uranium-plutonium cycle are presented in Fig. 1.

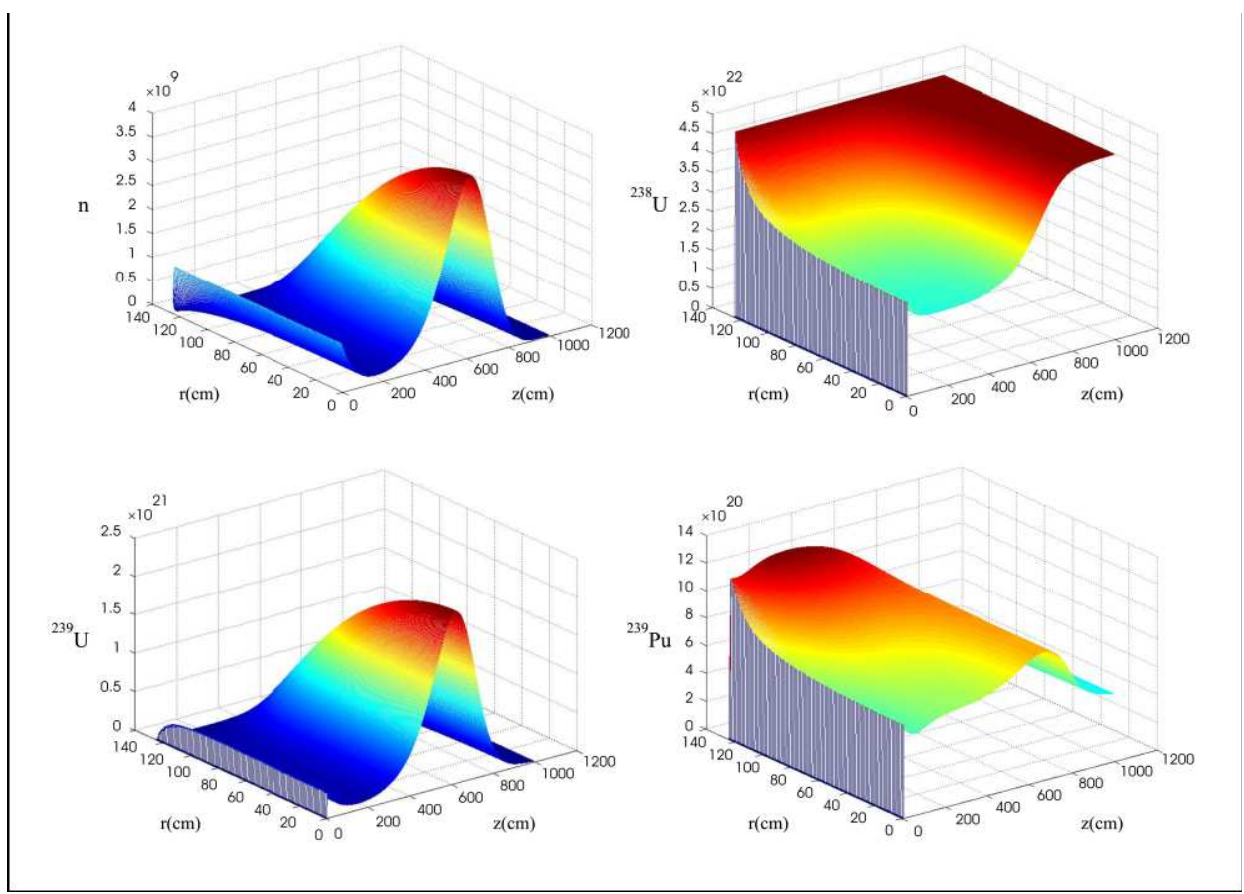

Fig. 1. Concentration kinetics of neutrons, $^{238}$U, $^{239}$U, $^{239}$Pu in the core of cylindrical reactor with radius of 125 cm and 1000 cm long at the time of 240 days. Here $r$ is transverse spatial coordinate axis (cylinder radius), $z$ is longitudinal spatial coordinate axis (cylinder length).

Within the framework of soliton-like fast reactor theory it is easily show that phase velocity $u$ of soliton-like neutron wave of the nuclear burning is generally predetermined by following approximate equality:

$$\frac{u\tau_\beta}{2L} \cong \left(\frac{8}{3\sqrt{\pi}}\right)^6 a^4 \exp\left(-\frac{64}{9\pi}a^2\right), \quad a^2 = \frac{\pi^2}{4} \cdot \frac{n_{crit}}{\tilde{n}_{fis} - n_{crit}}, \qquad (3)$$

where $\tilde{n}_{fis}$ and $n_{crit}$ are the equilibrium and critical concentrations of active (fissile) isotope, respectively; $L$ is the average diffusion distance of neutron, $\tau_\beta$ is delay time caused by active (fissile) isotope production, which amounts to the effective period of intermediate nuclei $\beta$-decay in uranium-plutonium cycle (1) or in thorium-uranium cycle (2).

Note that Eq. (3) automatically contains the self-regulation condition of nuclear burning because the wave existence is predetermined by inequality $\tilde{n}_{fis} > n_{crit}$. In other words Eq.(3) is the necessary physical requirement for the existence of the soliton-like neutron wave of nuclear burning. As it follows from Eq. (3), the upper bounds of the phase velocity of nuclear burning wave are 3.70 cm/day for uranium-plutonium cycle (1) and 0.31 cm/day for thorium-uranium cycle (2) at almost equal average diffusion distance ($L\sim5$ cm) of fast neutrons (1 MeV) both for uranium and thorium.

Finally, we consider the some important details and properties of such soliton-like fast reactor, assuming the existence of which, we obtain the theoretical spectra of reactor and terrestrial antineutrino which are in good agreement with the experimental KamLAND data [5] corresponding to first [1] and third [3] expositions.

According to our notions [5], a soliton-lke fast reactor is located on the boundary of the liquid and solid phases of the Earth. The average thickness of such a shell-boundary, which has increased density and mosaic structure, is ~2.2 km [11]. In our opinion, the most advanced mechanism for formation of such a shell below the mantle now are the experimental results of Anisichkin et al. [12]. According to these results, the chemically stable high-density actinide compounds (particularly uranium carbides and uranium dioxides) lose most of their lithophilic properties at high pressure, sink together with melted iron and concentrate in the Earth's core consequent to the initial gravitational differentiation of the planet. On the other words, during early stages of the evolution of the Earth and other planets, U and Th oxides and carbides (as the most dense, refractory, and marginally soluble at high pressures) accumulated from a magma "ocean" on the solid inner core of the planet, thereby activating chain nuclear reactions, and, in particular, a progressing wave of Feoktisov and/or Teller-Ishikawa-Wood type.

What is the thermal power of such a reactor? As a natural quantitative criterion of the georeactor thermal power we used the well-known (based on the geochemical measurements) $^3$He/$^4$He radial distribution in the Earth's interior [5]. It turned out that the experimental average values of $^3$He/$^4$He for crust, the depleted upper mantle, the mantle (minus the depleted upper mantle) and the so called $D''$-region in the lower mantle are in good agreement with the theoretical data obtained by the model of Feoktistov's uranium-plutonium georeactor with thermal power of 30 TW. Fig.2 shows the especial experimental investigation of geologically produced antineutrinos with KamLAND [3] and an alternative description of these data by georeactor model [5].

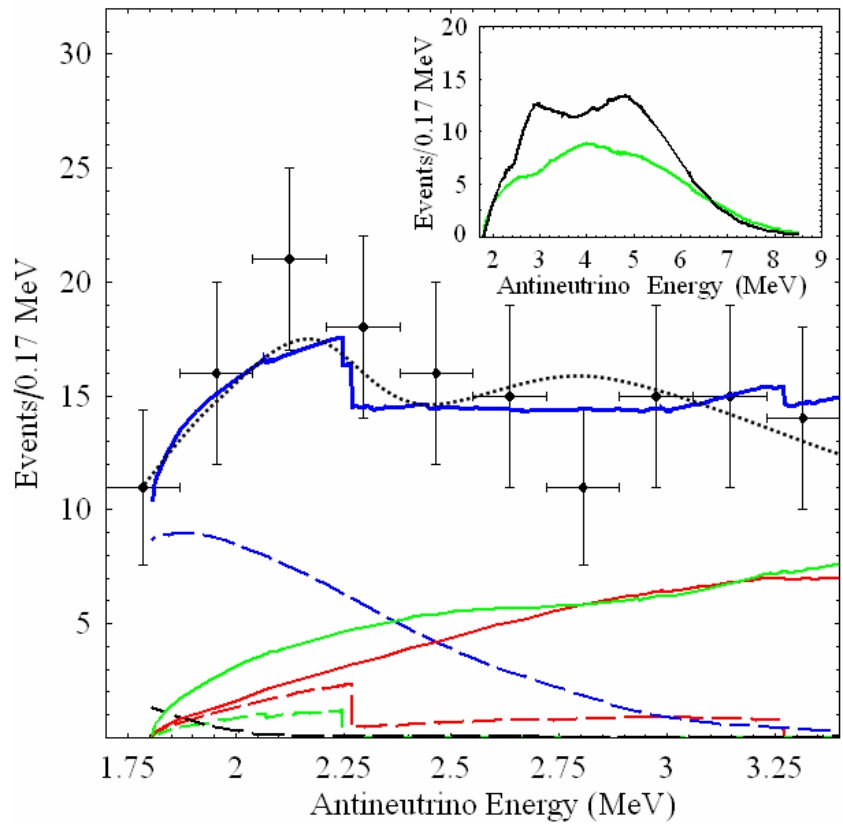

Fig. 2. The $\tilde{\nu}_e$ energy spectra in KamLAND [5]. Main panel, experimental points (solid black dots with error bars) together the total expectation obtained in KamLAND experiment (dotted black line) [3] and presented paper (thick solid blue line). Also shown are expected neutrino spectrum (solid green line) from Japan's reactor, the expected neutrino spectrum from georeactor of 30 TW(red line), the expected signals from $^{238}U$ (dashed red line) and $^{232}Th$ (dashed green line) geoneutrinos, $^{13}C(\alpha,n)^{16}O$ reactions (dashed blue line) and accidentals (dashed black line). Inset, expected spectra obtained in KamLAND experiment (solid black line) [3] and our paper [5] (solid green line) extended to higher energy.

Thereupon let us estimate the uncertainty of nuclear georeactor thermal power and then the uncertainty of the georeactor antineutrino spectrum. There is a need to notice here that, generally speaking, such uranium-plutonium georeactor can consist of a few hundreds or thousands reactors (with the total thermal power of 30 TW), which represents the individual burning "rivers' and 'lakes" of an inhomogeneous actinide shell located in the valleys of rough surface of the Earth's solid core [5]. In the general case, the fission rate of $^{239}Pu$ for the uranium-plutonium cycle (1) it is possible to write down in the form

$$\lambda_{Pu} = \phi \sigma_f n_{Pu} V \,, \tag{4}$$

where $\Phi = \upsilon n$ is the neutron-flux density; $\upsilon$ is the neutron velocity; $n$ is the neutron concentration; $\sigma_f$ is the fission cross-section of $^{239}Pu$; $n_{Pu}$ is $^{239}Pu$ concentration; $V$ is the volume of burning area.

It is easily seen, that due to the randomicity of the critical and equilibrium concentrations of plutonium in an actinide shell and also stochastic geometry of the "rivers" and "lakes" of actinide medium the relative variation of neutron flux density, plutonium concentration and size of burning areas can run up to 50% and more. In full measure it concerns the fission cross-section, whose variations are determined, first of all, by the non-trivial thermodynamics features (temperature and pressure) of actinide shell. In this case for the relative variation of fission velocity following relation is right:

$$\frac{\Delta \lambda_{Pu}}{\lambda_{Pu}} = \left[ \left( \frac{\Delta \phi}{\phi} \right)^2 + \left( \frac{\Delta n_{Pu}}{n_{Pu}} \right)^2 + \left( \frac{\Delta V}{V} \right)^2 \right]^{1/2} \geq 0.87 \,, \quad \frac{\Delta \sigma}{\sigma} << 1. \tag{5}$$

Let us show to what uncertainty of georeactor antineutrino oscillation spectrum the relative error of the fission rate of plutonium (5) leads. For this purpose we write down the theoretical form of measured total energy antineutrino oscillation spectrum $dn/dE$ in the $i$th energetic bin:

$$n_i(E) = m_\lambda \nu_i(E) \,, \tag{6}$$

where

$$m_\lambda = \lambda_{Pu} \Delta t, \quad \nu_i(E) = \frac{\varepsilon_i N_p}{4 \pi L^2} \sum_{j,i} \alpha_i \rho_{ji}(E) \sigma_{\nu p}(E) p(E, L) \,. \tag{7}$$

Here $m_\lambda$ is the total number of fissions during exposure time $\Delta t$ determined by fission rate $\lambda_{Pu}$; $\nu_i(E)$ is the average number of detected antineutrino per fission in the $i$th energetic bin; $\varepsilon$ is the detection efficiency of positrons of the inverse $\beta$-decay reaction; $N_p$ is the number of protons in the detector sensitive volume; $\Delta t$ is exposure time; $p(E,L)$ is neutrino oscillation probability with the corresponding parameters of mixing and energy $E$ at the distance $L$ from reactor; $(1/4\pi L^2)$ is effective solid angle; $\sigma_{vp}$ is the antineutrino-proton interaction cross-section of inverse $\beta$-decay reaction with the corresponding radiation corrections [5]; $\Sigma\alpha_i\rho_i(E)$ is the energy antineutrino spectrum of nuclear fuel in the $i$th energetic bin, MeV/fission; $\alpha_i$ is the part of $i$th isotope.

Here we should note that, in general, it is rightfully to consider the energy antineutrino spectra at different reactor heat power as self-similar, that considerably lightens its further analysis. At the same time the self-similarity takes place only for equilibrium neutrino spectra [13] typical for stationary processes in reactor core. And vice versa, the self-similarity conditions for equilibrium neutrino spectra are violated at nonstationary processes in the reactor core. This means, if, for example, the variations of energy neutron spectrum (and therefore the variations of mass yields induced by the fission of $^{239}Pu$) in the reactor core are considerable, corresponding neutrino spectra are not self-similar. Therefore calculated ("stationary") spectra and experimental ("nonstationary") spectra are differ up to 25% and higher [13].

Obviously, due to the stochastic change of nuclear georeactor heat power caused by the variations of neutron flux, plutonium concentration, medium geometry (5) and georeactor neutrino spectrum (7) the relative uncertainty of georeactor antineutrino oscillation spectrum $n_i^{geo}(E)$ in the $i$th energy bin (with allowance for Eqs.(5)-(7)) looks like:

$$\frac{\Delta n_i^{geo}}{n_i^{geo}} \cong \left[\left(\frac{\Delta\lambda_{Pu}}{\lambda_{Pu}}\right)^2 + \left(\frac{\Delta\rho_i}{\rho_i}\right)^2\right] \geq 0.9 \,, \qquad (8)$$

where $(\Delta\rho_i/\rho_i) \geq 25\%$ is the relative uncertainty of georeactor neutrino spectrum.

Therefore the lower estimation of the uncertainty of general antineutrino oscillation spectrum, which taking in account Eq.(8) and the contribution of uncertainty (4.14%) of antineutrino spectrum $n_i^{Jap}(E)$ from the Japanese reactors [4], become

$$\Delta n_i = \left[(0.0414 n_i^{Jap})^2 + (0.9 n_i^{geo})^2\right]^{1/2}; \qquad (9)$$

Note that just this uncertainty is shown as a violet band on the blue histogram in Fig.3.

Now we are ready to use the model of uranium-plutonium georeactor [5] for the alternative description of the data produced in new KamLAND experiment [4]. Obviously, that the standard methods of obtaining consistent estimates (e.g., the maximum-likelihood method) normally used for the determination of the oscillation parameters ($\Delta m_{12}^2$, $tg^2\theta_{12}$) [1-4] must take into account one more reactor, or, more specifically, account for the antineutrino spectrum of georeactor with the power ~ 30 TW which is located at a depth of $L$ ~ $5.2 \cdot 10^6$ m. However following [5], we apply here a simplified approach, the results of which application show that the hypothesis of the existence of a georeactor on the boundary of liquid and solid phases of the Earth's core does not conflict with the experimental data.

So, we proceed as in Ref. [5]: if CPT-invariance is assumed, the probabilities of the $\nu_e \to \nu_e$ and $\tilde{\nu}_e \to \tilde{\nu}_e$ oscillations should be equal at the same values $L/E$. On the other hand, it is known that the variations of $\Delta m^2$ are dominated over the more stable small variations of angle $\theta$ in the spectral distortion (oscillations) of the "solar" neutrino spectrum. Therefore we can assume (basing on CPT-theorem) that the angle which is determined by the experimental "solar" equality $\tan^2\theta_{12} =0.447$ [15] may be used as the reference angle of mixing in KamLAND-experiment.

Finally, following the computational ideology of Ref. [5] we give the results of the verification of optimal oscillation parameters ($\Delta m^2_{21} =2.5\cdot10^{-5}$ eV$^2$, $\tan^2\theta_{12} =0.437$) in the framework of the test problem of comparing the theoretical (which takes into account the georeactor operation) and the experimental spectrum of the reactor antineutrino on the base of new KamLAND data [4] (Fig.3). We compare also the $\chi^2$-profiles for KamLAND- hypothesis without georeactor and our georeactor hypothesis (Fig. 4).

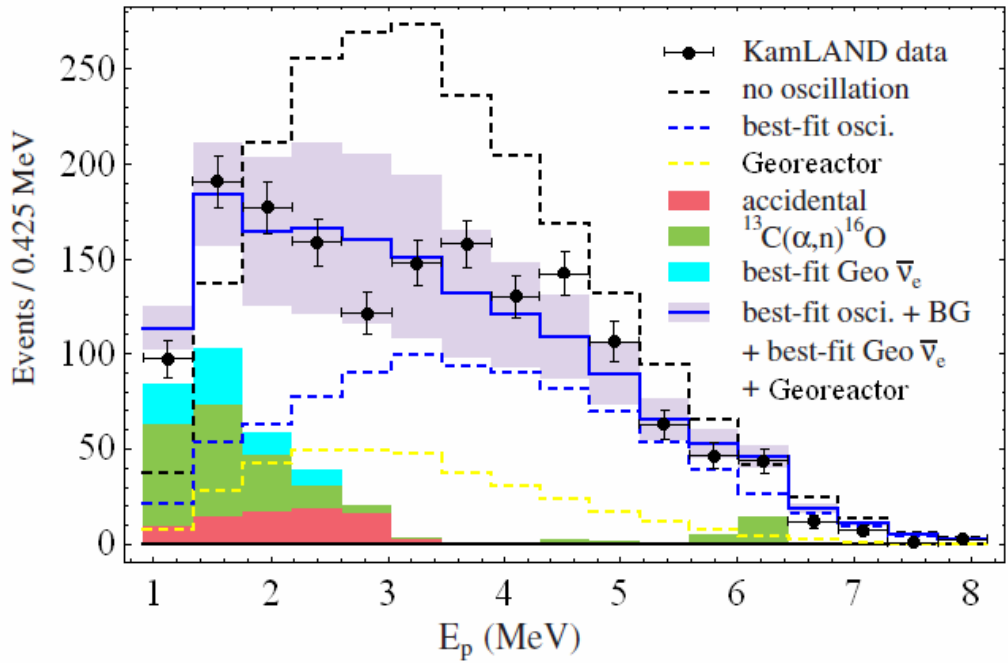

Fig. 3. Prompt event energy spectrum of $\tilde{\nu}_e$ candidate events. The shaded background and geoneutrino histograms are cumulative. Statistical uncertainties are shown for the data; the band on the blue histogram indicates the event rate systematic uncertainty. The georeactor power is 19.5 TW.

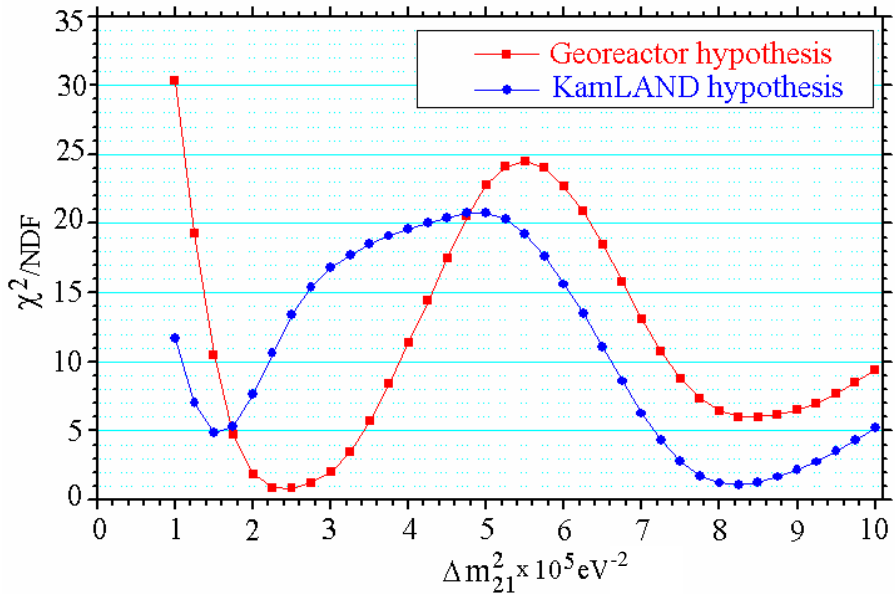

Fig. 4. Dependences of $\chi^2$/NDF on the mass squared difference $\Delta m_{21}^2$ corresponding to KamLAND- hypothesis without georeactor (blue line, tg$^2\theta_{12}$=0.56 [4]) and our georeactor hypothesis (red line, tg$^2\theta_{12}$=0.437).

Not looking at the low statistics of neutrino events (≤150 events/bin), the theoretical (which takes into account the soliton-like nuclear georeactor with power 19.5 TW) reactor antineutrino spectrum describes with acceptable accuracy the new experimental KamLAND-data [4] (Fig. 3). Here we pay attention to some important moments. Firstly, the average georeactor heat power is changed from 30 TW at the exposure time of 749.1±0.5 days in 2005 [3] (Fig. 2) to ~ 20 TW at total exposure of 1890 days in 2008 [4] (Fig.3). This reflects the nonstationary nature of the georeactor. Secondly, we have positive in respect to CPT-theorem physical fact that the parameters of mixing obtained within the framework of georeactor hypothesis are substantially closer to the data of solar flux SNO-experiment ($\Delta m^2{}_{21}$=4.57·10$^{-5}$ eV$^2$, tan$^2\theta_{12}$=0.447 [15]) then the parameters of mixing ($\Delta m^2{}_{21}$=7.58·10$^{-5}$ eV$^2$, tan$^2\theta_{12}$=0.56 [4]) obtained in KamLAND-experiment.

In conclusion, we should note that although the nuclear georeactor hypothesis which we used for the interpretation of KamLAND-experiment seems to be very effective, it can be considered only as a possible alternative variant for describing the KamLAND experimental data. In this sense, paradoxically as it may seem, only direct measurements of the geoantineutrino spectrum in the energy range >3.4 MeV in the future underground or submarine experiments will finally settle the problem of the existence of a natural georeactor and will make it possible to determine the "true" values of the reactor antineutrino oscillation parameters. At the same time just solution of the direct and the inverse problems of the remote neutrino tomography for the intra-terrestrial processes which is essential to obtain the pure geoantineutrino spectrum and to determine correctly the radial profile of the $\beta$–sources in the Earth's interior [14,16] will undoubtedly solve the problem of the existence of a natural nuclear reactor on the boundary of the liquid and solid phases of the Earth's core as well as the determination of the true geoantineutrino spectrum.